%% file: main.tex
\documentclass{article}
\usepackage{spconf}
\usepackage{amsmath}
\usepackage{amsfonts}
\usepackage{graphicx}
\usepackage{booktabs}
\usepackage{multirow}
\usepackage{caption}
\usepackage{subcaption}

\usepackage{pifont}
\newcommand{\cmark}{\ding{51}}
\newcommand{\xmark}{\ding{55}}

\title{A Comparison of Transformer, Convolutional, and Recurrent Neural Networks on Phoneme Recognition}
\name{Kyuhong Shim and Wonyong Sung}
\address{
Department of Electrical and Computer Engineering, Seoul National University, Korea\\
\small{\texttt{\{skhu20, wysung\}@snu.ac.kr}}
}

\begin{document}
\maketitle

% ---------------------------------------------------------------- %
\begin{abstract}
\input{sections/abstract}
\end{abstract}


% ---------------------------------------------------------------- %
\input{sections/introduction}
% ---------------------------------------------------------------- %
\input{sections/related}
% ---------------------------------------------------------------- %
\input{sections/model}
% ---------------------------------------------------------------- %
\input{sections/phoneme}
% ---------------------------------------------------------------- %
\input{sections/analysis}
% ---------------------------------------------------------------- %
\input{sections/conclusion}
% ---------------------------------------------------------------- %

% \section{ACKNOWLEDGMENTS}\label{sec:ack}
% Acknowledgements.

% ---------------------------------------------------------------- %
\newpage
\bibliographystyle{IEEEbib}
\bibliography{refs}

\end{document}

%% file: sections/abstract.tex
Phoneme recognition is a very important part of speech recognition that requires the ability to extract phonetic features from multiple frames.
In this paper, we compare and analyze CNN, RNN, Transformer, and Conformer models using phoneme recognition.
For CNN, the ContextNet model is used for the experiments.
First, we compare the accuracy of various architectures under different constraints, such as the receptive field length, parameter size, and layer depth.
Second, we interpret the performance difference of these models, especially when the observable sequence length varies. 
Our analyses show that Transformer and Conformer models benefit from the long-range accessibility of self-attention through input frames.

%% file: sections/introduction.tex
% ---------------------------------------------------------------- %
\section{Introduction}\label{sec:intro}
% ---------------------------------------------------------------- %

The ability to extract phonologically meaningful features is essential for various speech processing tasks such as automatic speech recognition (ASR)~\cite{deepspeech2,contextnet,conformer}, speaker verification~\cite{titanet}, and speech synthesis~\cite{neural_synthesis,glowtts}.
Such phoneme-awareness is a fundamental building block for human intelligence; not only the spoken but also the written language directly corresponds to the combination of phonemes.
% Recently, deep neural networks (DNNs) have achieved great success in speech applications over traditional handcrafted and rule-based techniques, thanks to the powerful ability to automatically capture phonetic characteristics.

In speech processing, DNN architectures can be categorized by how the feature extraction mechanism incorporates past and future information.
First, convolutional neural networks (CNNs) exploit the fixed-length convolution kernel to aggregate multiple frame information.
Because each frame can only access nearby frames within the kernel size in a convolutional layer, CNN models often stack multiple layers to capture long-range relationships.
Second, recurrent neural networks (RNNs) compress the entire past/future sequence into a single feature vector.
This compression enables RNN to utilize the entire sequence efficiently; however, RNN suffers from the loss of long-range information because the representation space is restricted to a single vector.
In contrast, Transformer-based models process the entire sequence simultaneously using the self-attention, where each frame directly accesses every other frame and adaptively determines their importance~\cite{transformer}.
In other words, Transformer-based models are more advantageous for long-range dependency modeling compared to CNN and RNN models.
For this reason, Transformer has become the universal choice for state-of-the-art speech processing in recent years.
However, phoneme recognition is considered a task that depends on a very short time interval of speech when compared to linguistic processing.
Many phonemes can be classified even with only one or a few frames of speech.
Thus, the phoneme classification efficacy of DNNs, especially Transformer-based ones that can process long-rage relationships, needs to be studied in detail.

In this paper, we compare four different DNN architectures for phoneme recognition.
Specifically, we compare CNN, RNN, Transformer, and Conformer~\cite{conformer} models under the same conditions.
Then, we analyze how different components and limitations of each architecture affect the performance.
We emphasize that phoneme recognition is the most suitable task for evaluating the phonetic feature extraction capability.
This is because other speech-related tasks usually require a model to encapsulate more information than phonetic knowledge in features.
% This is because other speech-related tasks usually require a model to learn more than phonetic knowledge.
For example, for end-to-end speech recognition, the model should utilize phonetic and linguistic information together to generate correct transcription~\cite{understanding}.
For speaker verification, speaker diarization, and speech synthesis, the model must consider the non-phonetic aspects of the speech, such as pitch, accent, speed, or loudness.
On the other hand, phoneme recognition performance can be easily measured by accuracy, and the result solely depends on the feature quality.

% Considering that each phoneme is uttered within a very short period, one may think that there would be little benefit to looking at the long range.
% However, we show that Transformer and Conformer can achieve additional gain from increasing the attention restriction range, while the improvement obtained by increasing convolution kernel size in CNN is limited.
% Our work establishes a connection to recent studies on ASR models~\cite{understanding,similarity_content} and self-supervised speech models~\cite{wav2vec2,mockingjay,wav2vec-u} demonstrating that self-attention is capable of extracting rich phonetic features by learning diverse phonological relationships between frames.

We summarize our findings below:
\begin{itemize}
    \item Although each phoneme is uttered within a short period, the phoneme recognition accuracy of DNN is improved until the receptive field length is fairly long.
    \item When the receptive field length becomes longer, Transformer and Conformer show consistent performance improvement, in contrast to CNN.
    \item When the parameter size is very small, such as 1M, the ContextNet performs best. Also, ContextNet is advantageous when considering the inference time in GPU.
    % \item Conformer shows the best accuracy and is the most parameter-efficient architecture except when the parameter budget is very small.
    % \item Conformer shows the best accuracy in most configurations, while LSTM is the worst in every case.
    % \item Especially, Transformer and Conformer show more improvement in challenging utterances, such as LibriSpeech \textit{test-other}, than the clean one, when the receptive field length increases.
    % \item Conformer is the most parameter-efficient architecture except when the parameter budget is very small.
\end{itemize}

%% file: sections/related.tex
% ---------------------------------------------------------------- %
\section{Related Work}\label{sec:related}
% ---------------------------------------------------------------- %

% ---------------------------------------------------------------- %
\subsection{Phoneme recognition}
% ---------------------------------------------------------------- %

Earlier studies have first introduced neural networks for phoneme recognition~\cite{hmm_phoneme}, such as time-delay networks~\cite{tdnn} and bidirectional LSTM~\cite{lstm_phoneme}.
In these works, the benefit of considering more than about 10 frames was marginal.

Recently, phoneme recognition is widely used as a tool to evaluate the amount of phonetic information of DNN features learned from other tasks, including ASR and self-supervised learning.
For example, Mockingjay~\cite{mockingjay}, wav2vec 2.0~\cite{wav2vec2} and wav2vec-U~\cite{wav2vec-u} exploit phoneme recognition on self-supervised pre-trained Transformer models to demonstrate that their models learn general speech representations.
Our work is different from these works in that we directly train models on the phoneme recognition task.
By doing so, the model can fully utilize its capability in extracting phonetic characteristics without being distracted by other objectives.

% ---------------------------------------------------------------- %
\subsection{Transformer-based speech processing}
% ---------------------------------------------------------------- %

Several studies have investigated the behavior of Transformer models in order to understand their superior performance.
Probing experiments on the self-supervised Transformer models discovered that Transformers detect diverse aspects of audio, including voice pitch, fluency, duration, and phonemes~\cite{probing,probing2,layerwise}.
On the other hand, analyses on the attention map revealed that Transformer considers the entire sequence in phonetic feature extraction, named phonetic localization~\cite{understanding}.
For example, a self-attention head that performs phonetic localization would pay high attention weight for similarly pronounced frames.
% , regardless of the distance between frames.

Furthermore, different self-attention heads are specified for different phonetic relationships~\cite{understanding,yang20i}.
Specifically, phonetic self-attention behavior can be separated into similarity-based and content-based ones, where the former focuses on the pairwise similarity of frames while the latter considers the content of each frame~\cite{similarity_content}.
We note that such unique behaviors have not been reported in CNN- and RNN-based models.

% ---------------------------------------------------------------- %
\subsection{Comparison between Transformer and Others}
% ---------------------------------------------------------------- %

Extensive studies have been conducted to compare CNN to Transformers in the vision domain~\cite{vit_survey}.
Especially, comparisons between vision Transformer (ViT) and CNN show that they learn very different aspects of an image~\cite{do_vit}; for example, ViT and CNN behave as low-pass and high-pass filters, respectively~\cite{how_vit}.
However, in-depth analyses were not conducted much in the speech domain.
Several works have investigated RNN-based and Transformer-based models for ASR tasks~\cite{comparative,comparison,li20}, but only the final word error rate and training dynamics are compared.
In our experiments, we carefully design model configurations for a fair comparison and compare four architectures with the same constraints.

%% file: sections/model.tex
% ---------------------------------------------------------------- %
\section{DNN Architecture}\label{sec:model}
% ---------------------------------------------------------------- %

% In this section, we briefly introduce our target architectures: CNN, RNN, Transformer~\cite{transformer}, and Conformer~\cite{conformer}.

\subsection{CNN}\label{ssec:contextnet}
\begin{figure}[t]
    \centering
    \includegraphics[width=1.0\linewidth]{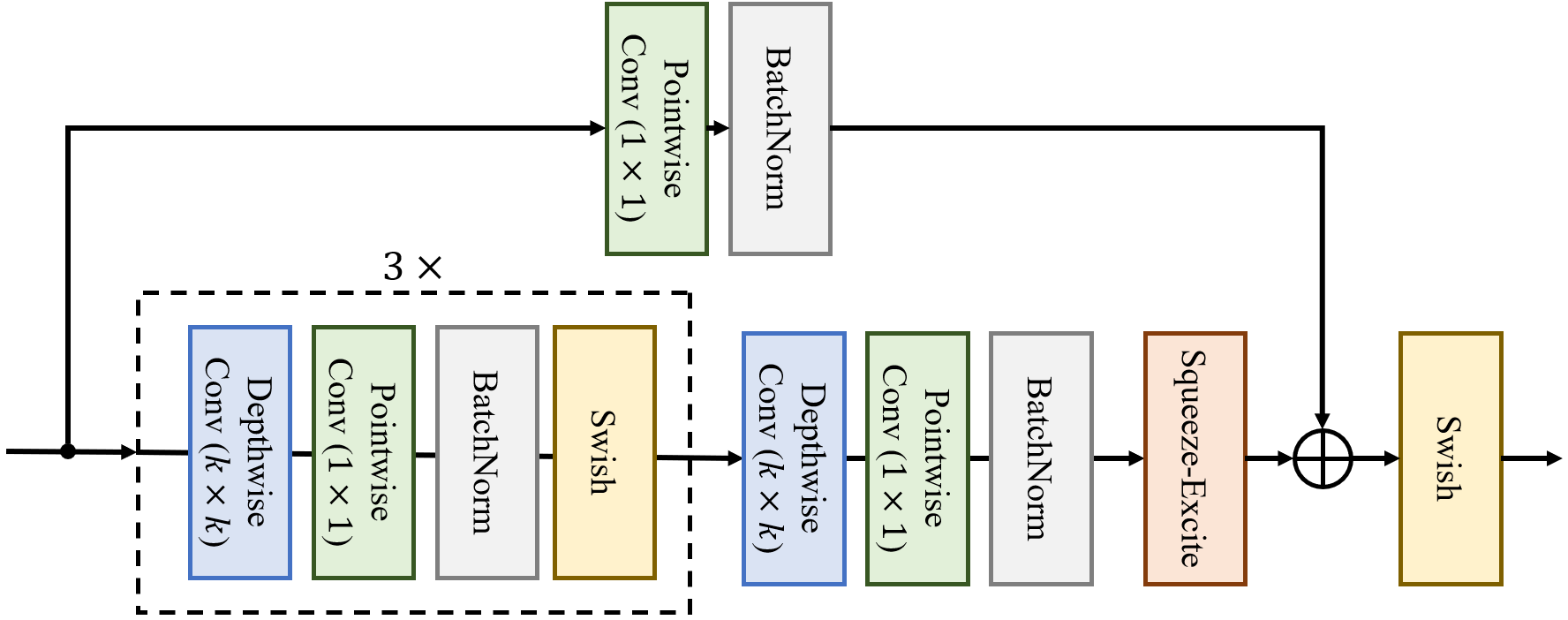}
    \caption{Illustration of a single ContextNet block.}
    \label{fig:contextnet}
\end{figure}

We choose ContextNet~\cite{contextnet} as a baseline because the ContextNet block has been employed in many state-of-the-art CNN-based ASR models~\cite{contextnet,quartznet}.
ContextNet architecture differs from other CNNs in two components: depthwise separable (DS) convolution~~\cite{mobilenetv2,contextnet,quartznet,tds} and squeeze-excite (SE) module~\cite{se}.
Figure~\ref{fig:contextnet} shows one ContextNet block that includes four DS convolution layers, residual connection, and SE module.
Note that we take a block as the basic unit of ContextNet for experiments.

First, DS convolution includes a depthwise convolution of large kernel size $k$ followed by a pointwise convolution of kernel size 1.
The former aggregates neighboring frames without mixing channels, and the latter combines every channel for each frame.
This two-step process makes DS convolution parameter-efficient because the model can increase the kernel size without increasing the number of parameters much.
Second, SE module adaptively re-weights channels based on the per-channel feature averaged through the entire sequence.
SE module is an efficient approach for incorporating the global information in feature processing, however, it does not consider the difference of frames because the same channel weights are multiplied by every frame feature.

% \begin{figure}[t]
%     \centering
%     \includegraphics[width=0.95\linewidth]{figures/conformer.png}
%     \caption{Illustration of a single Conformer layer.}
%     \label{fig:conformer}
%     % \vspace{-0.1cm}
% \end{figure}

\subsection{RNN}
We use LSTM~\cite{lstm} as our default RNN layer.
Specifically, we stack multiple bidirectional LSTM layers to build an RNN model~\cite{deepspeech2}.
Unlike other architectures, RNN-based models require sequential processing of frames, which causes a slow inference especially for bidirectional ones.

\subsection{Transformer}
We employ the pre-norm Transformer layer~\cite{prenorm} which includes two submodules: multi-head self-attention and feed-forward.
Please refer to the original work~\cite{transformer} for the internal structure of submodules.
While the post-norm design was employed in the original Transformer model, the pre-norm design is adopted in many speech and language processing models~\cite{conformer,trans_transducer}.

\subsection{Conformer}
We utilize the Conformer layer~\cite{conformer}, which is widely adopted in recent state-of-the-art ASR models~\cite{pushing,speechstew}.
The key idea of Conformer is to incorporate an additional convolution module between the self-attention and feed-forward submodules.
Due to this convolution module, Conformer can enhance the locality in extracted features.
% Figure~\ref{fig:conformer} visualizes the structure of a single Conformer layer.

%% file: sections/phoneme.tex
% ---------------------------------------------------------------- %
\section{Phoneme Recognition}\label{sec:phoneme}
% ---------------------------------------------------------------- %

\begin{figure}[t]
    \centering
    \includegraphics[width=0.7\linewidth]{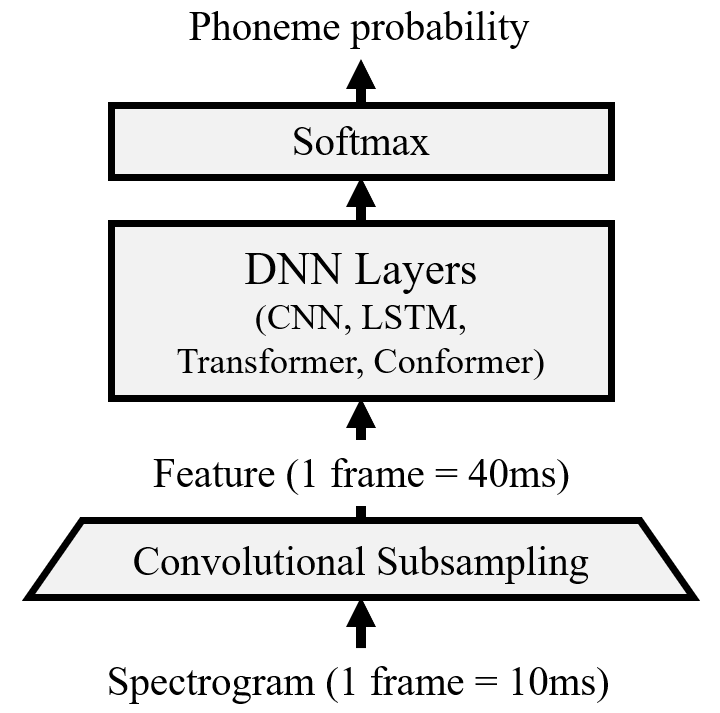}
    \caption{Basic DNN architecture for phoneme recognition.}
    \label{fig:architecture}
    % \vspace{-0.1cm}
\end{figure}

In this section, we compare various models from three perspectives: 1) receptive field length, 2) parameter size, and 3) layer depth.
For a fair comparison, we change each aspect while preserving the others.
% We train models on the phoneme recognition task and show their ability to extract phonetic features.

First, we vary the receptive field length to measure information efficiency.
The receptive field length represents how many frames incorporate in the feature extraction process.
Note that ContextNet with SE module~\cite{se} and LSTM cannot restrict the receptive field since these modules are designed to observe the entire sequence.
Second, we change the number of layers to investigate the trade-off between the depth and hidden dimension, or the depth and receptive field length, while maintaining the parameter size.
Finally, we compare different models with the same number of parameters to measure parameter efficiency.
% The number of parameters can also be a good proxy of the model capacity [ ].
% The number of parameters directly shows the required storage size, which can be a crucial factor for embedded devices.

% ---------------------------------------------------------------- %
\subsection{Setup}
% ---------------------------------------------------------------- %

We train the models on the LibriSpeech-960~\cite{librispeech} dataset using the frame-level phoneme alignment extracted from Montreal Forced Aligner~\cite{mfa}.
% Table~\ref{tab:phoneme_classes} shows the 36 phoneme classes of the datas
We use the 37 phoneme classes taken from the previous study~\cite{understanding}, including `silence'.
We utilize 80-dimensional Mel filterbank features as the input, extracted by 25ms window and 10ms stride.
 
Figure~\ref{fig:architecture} illustrates the common architecture for models used in phoneme recognition experiments.
The input features first pass through two convolutional subsampling layers~\cite{deepspeech2} of kernel size 3 and stride 2, in order to efficiently reduce the number of frames without much information loss.
% Note that the model outputs the class probability distribution of each frame among 37 classes including `silence'.
For the loss function, a simple per-frame cross-entropy is used.

The models are trained using AdamW~\cite{adamw} optimizer for 25K iterations with the maximum learning rate of 0.001, weight decay of 0.001, and cosine learning rate decay schedule.
The batch size is set to 128, and SpecAugment~\cite{specaugment} is employed for regularization.
The same training configuration is applied for every experiment.
% A single A100 (40GB) GPU is utilized, where each experiment takes about 5 to 8 hours depends on the configuration.

% \begin{table}[t]
%     \centering
%     \caption{List of phoneme classes for phoneme recognition. The order of phonemes follow the previous study~\cite{understanding}.}
%     \vspace{0.1cm}
%     \resizebox{1.0\linewidth}{!}{
%     \begin{tabular}{c|ccccc cccc}
%         \toprule
%          Cls. & 0 & 1 & 2 & 3 & 4 & 5 & 6 & 7 & 8 \\
%          Phn. & AA & AE & AW & AY & AH & EH & ER & EY & IY \\
%          \midrule
%          Cls. & 9 & 10 & 11 & 12 & 13 & 14 & 15 & 16 & 17 \\
%          Phn. & IH & O & UH & UW & L & R & M & N & NG \\
%          \midrule
%          Cls. & 18 & 19 & 20 & 21 & 22 & 23 & 24 & 25 & 26 \\
%          Phn. & B & D & DH & G & K & P & T & F & CH \\
%          \midrule
%          Cls. & 27 & 28 & 29 & 30 & 31 & 32 & 33 & 34 & 35 \\
%          Phn. & SH & TH & S & Z & V & JH & W & Y & HH \\
%          \bottomrule
%     \end{tabular}}
%     \label{tab:phoneme_classes}
% \end{table}

% ---------------------------------------------------------------- %
\subsection{Effect of the Receptive Field Length}
% ---------------------------------------------------------------- %

To investigate the effect of the receptive field length, we fix the number of parameters to 5M and set the number of layers to four.
For ContextNet, we either 1) set the number of blocks to four and change the convolution kernel size, or 2) stack more blocks without increasing the kernel size.
For Transformer and Conformer, we manually restrict the accessible self-attention range by masking the attention map.
For example, the receptive field length of CNN with four DS convolutional layers (i.e., a single ContextNet block) of kernel size $k$ is $4k-3$, which can be matched by restricting the self-attention range to $[-r, r] = [-2k+2, 2k-2]$.

\begin{table}[t]
    \centering
    \caption{Phoneme recognition accuracy (\%) of LSTM, Transformer, and Conformer models with different receptive field lengths. The parameter size is limited to about 5M, and the layer depth is set to 4. $k$, and $r$ indicate the convolution kernel size and the self-attention range restriction, respectively.
    }
    \vspace{0.1cm}
    \begin{tabular}{lr|cc}
        \toprule
        \multirow{2}{*}{Model}  & Receptive   & \textit{test-}  & \textit{test-} \\ 
                                & field (sec) & \textit{clean}  & \textit{other} \\
        \midrule
        LSTM                    & unlimited  & 87.4  & 79.6 \\  % 5,046,870
        \midrule
        Trans($r=8$)            &  2.60      & 87.5  & 79.9 \\
        Trans($r=16$)           &  5.16      & 87.7  & 80.1 \\
        Trans($r=32$)           & 10.28      & 87.9  & 80.5 \\
        Trans($r=64$)           & 20.52      & 88.0  & 80.8 \\
        Trans                   & unlimited  & 88.4  & 81.4 \\  % 5,086,206
        \midrule
        Conf($k=9,r=4$)         &  2.60      & 89.5  & 82.6 \\
        Conf($k=9,r=12$)        &  5.16      & 89.5  & 83.0 \\
        Conf($k=9,r=28$)        & 10.28      & 89.7  & 83.5 \\
        Conf($k=9,r=60$)        & 20.52      & 89.8  & 83.8 \\
        Conf($k=9$)             & unlimited  & \textbf{90.3}  & \textbf{84.4} \\  % 5,153,382
        \bottomrule
    \end{tabular}
    \label{tab:receptive_field}
    % \vspace{0.1cm}
\end{table}

\begin{table}[t]
    \centering
    \caption{Phoneme recognition accuracy (\%) of ContextNet models with different receptive field lengths and layer depths. The parameter size is about 5M for all models. $k$ and $l$ are the convolution kernel size and the number of stacked ContextNet blocks, respectively,
    ` * ' implies that the receptive field length is actually unlimited because of the SE module.
    }
    \vspace{0.1cm}
    \begin{tabular}{lr|cc}
        \toprule
        \multirow{2}{*}{Model}  & Receptive   & \textit{test-}  & \textit{test-} \\ 
                                & field (sec) & \textit{clean}  & \textit{other} \\
        \midrule
        Context($k=3, l=4$)       &  1.32*      & 89.5  & 83.3 \\  % 5,049,334
        Context($k=5, l=4$)       &  2.60*      & 89.7  & 83.8 \\  % 5,060,598
        Context($k=9, l=4$)       &  5.16*      & 89.7  & 83.7 \\  % 5,083,126
        Context($k=17, l=4$)      & 10.28*      & 89.6  & 83.4 \\  % 5,128,182
        Context($k=33, l=4$)      & 20.52*      & 89.4  & 83.0 \\  % 5,218,294
        \midrule
        Context/SE($k=3, l=4$)       &  1.32      & 89.1  & 82.1 \\  % 4,923,846
        Context/SE($k=5, l=4$)       &  2.60      & 89.4  & 82.7 \\  % 4,935,110
        Context/SE($k=9, l=4$)       &  5.16      & 89.5  & 82.8 \\  % 4,957,638
        Context/SE($k=17, l=4$)      & 10.28      & 89.4  & 82.8 \\  % 5,002,694
        Context/SE($k=33, l=4$)      & 20.52      & 89.4  & 82.7 \\  % 5,092,806
        \midrule
        Context/SE($k=5, l=4$)  &  2.60      & 89.4  & 82.7 \\  % 4,935,110 (d=352)
        Context/SE($k=5, l=8$)  &  5.16      & 89.8  & 83.3 \\  % 5,029,782 (d=272)
        Context/SE($k=5, l=12$) & 10.28      & 89.7  & 83.3 \\  % 5,114,174 (d=232)
        Context/SE($k=5, l=16$) & 20.52      & 89.6  & 83.2 \\  % 4,933,790 (d=200)
        \bottomrule
    \end{tabular}
    \label{tab:receptive_field_cnn}
    \vspace{-0.1cm}
\end{table}

Table~\ref{tab:receptive_field} and ~\ref{tab:receptive_field_cnn} show the accuracy of four different DNN architectures.
`ContextNet/SE' indicates that the SE module is removed to prevent incorporating the global information of the entire sequence.
To equally set the parameter size, the hidden dimension of ContextNet($l=4$), LSTM, Transformer, and Conformer is set to 352, 336, 248, and 192, respectively.
We find out several interesting results:
\begin{itemize}
    \item LSTM shows the worst performance although its theoretical receptive field length is unlimited. This observation is consistent with every experiment.
    \item Transformer and Conformer benefit from increasing the receptive field length, but the accuracy of ContextNet saturates from a certain point. The best performance for CNN models is achieved with a receptive field length of 5.16 seconds.
    \item Transformer with unlimited receptive field length is worse than Conformer with the shortest length, emphasizing the importance of convolutional layers in phonetic feature extraction.
\end{itemize}
% The findings emphasize the importance of convolutional layers in phonetic feature extraction; ContextNet and Conformer present consistently better performance than LSTM and Transformer.
In fact, convolution and self-attention extract phonetic information differently by focusing on local and global regions of the sequence, respectively.
A recent study discovered a behavior called \textit{phonetic localization}~\cite{understanding} that the self-attention module captures the phonetic relationship through the entire sequence.
For example, in self-attention, a frame tends to pay larger attention weights to similar phonemes (e.g., `TH' to 'SH', `N' to `M', `S' to `Z').
Similar frames affect each other in the feature domain, and as a result, they become more clustered based on phoneme classes.

Considering that the same phoneme can be pronounced differently, it is advantageous to standardize features of the same class for phoneme recognition.
The essence of phonetic localization is that every frame can access all frames dynamically.
When similar frames appear in the sequence broadly, self-attention can aggregate such related information regardless of the distance between frames, while convolution cannot utilize the phonetic connection between far frames.
We believe the phonetic localization is the reason why the accuracy improves as the receptive field length increases for models that equip self-attention.

% ---------------------------------------------------------------- %
\subsection{Effect of the Number of Layers}
% ---------------------------------------------------------------- %

It is known that utilizing multiple layers helps extract more complex and rich features.
On the other hand, a larger hidden dimension is often considered to provide better expressiveness for each frame feature.
Therefore, increasing the width(hidden dimension) or the depth(the number of layers) is an interesting design choice.
In this subsection, we increase the number of layers while preserving the parameter size (5M) and receptive field length.
Specifically, we decrease the hidden dimension as the number of layers increases.
% For convolutional layers, we additionally adjust the kernel size to maintain the receptive field length.

\begin{table}[t]
    \centering
    \caption{Phoneme recognition accuracy (\%) of DNN models with different number of layers.}
    \vspace{0.1cm}
    \begin{tabular}{lr|cc}
        \toprule
        \multirow{2}{*}{Model}  & \multirow{2}{*}{\#Layer}   & \textit{test-}  & \textit{test-} \\ 
                                &                           & \textit{clean}  & \textit{other} \\
        % \midrule
        % Context($d=448, k=9$) & 2   & 89.0  & 82.4 \\  % 5,043,286
        % Context($d=352, k=5$) & 4   & 89.7  & 83.8 \\  % 5,060,598
        % Context($d=304, k=3$) & 6   & 89.9  & 84.0 \\  % 5,122,106
        % Context($d=264, k=3$) & 8   & 89.9  & 84.1 \\  % 4,939,686
        \midrule
        LSTM($d=432$)            & 2   & 83.3  & 74.0 \\  % 5,068,278
        LSTM($d=336$)            & 4   & 87.4  & 79.6 \\  % 5,046,870
        LSTM($d=288$)            & 6   & 88.1  & 80.7 \\  % 5,078,790
        LSTM($d=256$)            & 8   & 88.1  & 80.9 \\  % 5,076,006
        \midrule
        Trans($d=328$)          & 2   & 85.9  & 77.8 \\  % 5,093,022
        Trans($d=248$)          & 4   & 88.4  & 81.4 \\  % 5,086,206
        Trans($d=208$)          & 6   & 89.3  & 82.7 \\  % 5,059,990
        Trans($d=184$)          & 8   & 89.7  & 83.6 \\  % 5,085,918
        \midrule
        Conf($d=256, k=17$)     & 2   & 88.9  & 82.2 \\  % 5.084,198
        Conf($d=192, k=9$)      & 4   & 90.3  & 84.4 \\  % 5,153,382
        Conf($d=160, k=7$)      & 6   & 90.6  & 85.0 \\  % 5,141,510
        Conf($d=140, k=5$)      & 8   & \textbf{90.9}  & \textbf{85.3} \\  % 5,118,970
        \bottomrule
    \end{tabular}
    \label{tab:layers_depth}
    \vspace{-0.1cm}
\end{table}

% Table~\ref{tab:receptive_field_cnn} and ~\ref{tab:layers_depth} present the effect of the number of layers for different architectures.
The effect of the layer depth for ContextNet/SE is shown in Table~\ref{tab:receptive_field_cnn} and that for the RNN, Transformer, and Conformer is presented in Table~\ref{tab:layers_depth}.
Since one ContextNet/SE layer contains four DS convolution layers, the ContextNet/SE with $l=16$ actually connects 64 DS convolution layers in series.
As expected, utilizing more layers improves the accuracy without exception.
We summarize our findings as below:
\begin{itemize}
    \item The benefit of increasing the number of layers in Transformer and Conformer is higher than in the other two architectures.
    % CNN and LSTM models show saturation in accuracy when more than 6 layers are used.
    \item Conformer achieves the highest parameter efficiency when using more than 4 layers. For example, the 4-layer Conformer model already achieves higher accuracy than other 8-layer models.
    % \item Increasing the number of layers is more beneficial than increasing the hidden dimension.
\end{itemize}

% ---------------------------------------------------------------- %
\subsection{Effect of the Parameter Size}
% ---------------------------------------------------------------- %

To evaluate the relationship between the parameter size and phoneme feature extraction, we change the hidden dimension $d$ of each model and compare the accuracy.
For ContextNet and Conformer, the convolution kernel size is set to $k=9$ as in Table~\ref{tab:receptive_field}.
Note that ContextNet models in this experiment observe the entire sequence by utilizing the SE module.
% Basically, we do not limit the receptive field length, but for CNN, we use the receptive field length of about 2.6 seconds because it shows the best accuracy (see CNN($k=9$) in Table~\ref{tab:receptive_field}).

\begin{table}[t]
    \centering
    \caption{Phoneme recognition accuracy (\%) of DNN models with different parameter sizes. The layer depth is set to two, and every model utilizes unlimited receptive field length.}
    \vspace{0.1cm}
    \begin{tabular}{lr|cc}
        \toprule
        \multirow{2}{*}{Model}  & Parameter   & \textit{test-}  & \textit{test-} \\ 
                                & size (M)    & \textit{clean}  & \textit{other} \\
        \midrule
        Context($d=248$)     & 1.04   & 86.8  & 79.0 \\  % 1,035,996
        Context($d=416$)     & 3.09   & 88.6  & 81.9 \\  % 3,089,996
        Context($d=448$)     & 5.04   & 89.0  & 82.4 \\  % 5,043,286
        \midrule
        LSTM($d=240$)           & 1.05   & 82.6  & 73.2 \\  % 1,049,334
        LSTM($d=392$)           & 3.02   & 83.4  & 74.2 \\  % 3,018,206
        LSTM($d=432$)           & 5.07   & 83.3  & 74.0 \\  % 5,068,278
        \midrule
        Trans($d=168$)          & 1.00   & 83.7  & 75.2 \\  % 998,526
        Trans($d=288$)          & 3.06   & 85.7  & 77.6 \\  % 3,063,430
        Trans($d=328$)          & 5.09   & 85.9  & 77.8 \\  % 5,093,022
        \midrule
        Conf($d=128$)           & 1.01   & 86.1  & 78.6 \\  % 1,005,286
        Conf($d=216$)           & 2.97   & 88.5  & 81.7 \\  % 2,970,766
        Conf($d=256$)           & 5.08   & 88.9  & 82.2 \\  % 5,084,198
        \bottomrule
    \end{tabular}
    \label{tab:param_size}
    % \vspace{0.1cm}
\end{table}

Table~\ref{tab:param_size} compares DNN architectures with the same parameter budget.
In particular, we evaluate the models with a small parameter size to assume an environment with very limited resources.
For 1M, 3M, and 5M size models, the number of channels in the convolutional subsampling (see Figure~\ref{fig:architecture}) is set to 64, 128, and 256, respectively.
% Noticeably, the hidden dimension of CNN models is the largest among architectures, thanks to the time-depth separable convolution.
% In contrast, the hidden dimension of Conformer models is the smallest because Conformer contains two feed-forward modules in a single layer.
% In all parameter size budgets, the order of phoneme recognition accuracy is LSTM $<$ Transformer $<$ Conformer $<$ CNN.
We observe that ContextNet and LSTM are the best and the worst parameter-efficient architecture in this small parameter size configuration with two layers.
However, when the layer depth is 4 (see Table~\ref{tab:receptive_field}), Conformer achieves higher accuracy than ContextNet with the same receptive field length and parameter size.
This implies that the advantage of ContextNet in parameter efficiency may disappear for modern DNN architectures that stack many layers.
% Note that the results in Table~\ref{tab:param_size} are obtained with 2-layer models.

%% file: sections/analysis.tex
% ---------------------------------------------------------------- %
\section{Analysis}\label{sec:analysis}
% ---------------------------------------------------------------- %

% In this section, we analyze the phonetic feature extraction mechanism of each architecture.
% In particular, we focus on two architectures, ContextNet and Conformer, which present a considerable accuracy gap over LSTM and Transformer.

% ---------------------------------------------------------------- %
\subsection{Depth-separable and Squeeze-Excite}\label{ssec:cnn_ablation}
% ---------------------------------------------------------------- %

As explained in Section~\ref{ssec:contextnet}, ContextNet exploits two specially designed components, DS convolution and SE module.
We conduct an ablation study on these modules to evaluate the effect of each ingredient in phonetic feature extraction.
When removing the DS convolution, we replace the cascading two convolution layers, depthwise and pointwise convolutions (see Figure~\ref{fig:contextnet}), with a single full convolution of kernel size $k$.
Because the full convolution equips more parameters, we reduce the hidden dimension to preserve the number of parameters.

% \begin{table}[t]
%     \centering
%     \caption{Ablation of depth separable (DS) convolution and SE module for phoneme recognition.}
%     \vspace{0.2cm}
%     \begin{tabular}{lccc|cc}
%         \toprule
%         \multirow{2}{*}{Model} & \multirow{2}{*}{DS} & \multirow{2}{*}{SE} & \multirow{2}{*}{Params} & \textit{test-} & \textit{test-} \\
%          & & & & \textit{clean} & \textit{other} \\
%         \midrule
%         CNN($d=448$)      & \cmark & \cmark &  5M  & \textbf{89.0}    & \textbf{82.4} \\  % 5,043,286
%         CNN($d=448$)      & \cmark & \xmark &  5M  & 88.8    & 81.7 \\  % 5,070,654 actually 456
%         CNN($d=212$)      & \xmark & \cmark &  5M  & 88.4    & 81.3 \\  % 5,039,518
%         CNN($d=212$)      & \xmark & \xmark &  5M  & 88.1    & 80.9 \\  % 5,016,994
%         \midrule
%         CNN($d=448$)      & \xmark & \cmark & 18M  & 90.0    & 84.0 \\  % 17,866,838
%         CNN($d=448$)      & \xmark & \xmark & 18M  & 89.9    & 83.4 \\  % 17,765,478
%         \bottomrule
%     \end{tabular}
%     \label{tab:cnn_ablation}
%     % \vspace{0.1cm}
% \end{table}

\begin{table}[t]
    \centering
    \caption{Ablation of depth separable (DS) convolution and SE module for phoneme recognition. The layer depth is fixed to two.}
    \vspace{0.1cm}
    \begin{tabular}{lccc|cc}
        \toprule
        \multirow{2}{*}{Model} & \multirow{2}{*}{DS} & \multirow{2}{*}{SE} & \multirow{2}{*}{Params} & \textit{test-} & \textit{test-} \\
         & & & & \textit{clean} & \textit{other} \\
        \midrule
        Context($d=352$)      & \cmark & \cmark &  5M  & \textbf{89.7}    & \textbf{83.8} \\  % 5,060,598
        Context($d=352$)      & \cmark & \xmark &  5M  & 89.4    & 82.7 \\  % 4,935,110
        Context($d=200$)      & \xmark & \cmark &  5M  & 89.4    & 83.1 \\  % 5,033,890
        Context($d=200$)      & \xmark & \xmark &  5M  & 89.1    & 82.1 \\  % 4,992,990
        \midrule
        Context($d=352$)      & \xmark & \cmark & 13M  & 90.6    & 85.0 \\  % 12,956,662
        Context($d=352$)      & \xmark & \xmark & 13M  & 90.3    & 84.1 \\  % 12,831,174
        \bottomrule
    \end{tabular}
    \label{tab:cnn_ablation}
    % \vspace{0.1cm}
\end{table}

Table~\ref{tab:cnn_ablation} demonstrates the importance of the DS convolution and SE module for ContextNet, using a 5M parameter size budget and a layer depth of two.
In summary, the combination of DS convolution and SE module is the most parameter-efficient configuration.
In addition, the accuracy loss caused by replacing DS convolution with full convolution ($-0.7\%$) is smaller than removing the SE module ($-1.1\%$) in \textit{test-other} dataset.
When the hidden dimension is not reduced, the model achieves the best accuracy as expected, at a cost of about 2.6 times more parameters.

% ---------------------------------------------------------------- %
\subsection{The Number of Self-Attention Heads}\label{ssec:sa_ablation}
% ---------------------------------------------------------------- %

Increasing the layer depth improves the accuracy (see Table~\ref{tab:layers_depth}), but this also increases the total number of (self-attention) heads.
To clarify the source of improvement, we conduct an ablation study on the number of heads; we vary the number of heads to 2, 4, 8, and 16.
Note that changing the number of self-attention heads in a Transformer and Conformer layer does not affect the number of parameters because the per-head hidden dimension changes accordingly.
Transformer and Conformer models in previous experiments employ four heads for every layer regardless of the hidden dimension.

\begin{table}[t]
    \centering
    \caption{Ablation on the number of self-attention heads for phoneme recognition.}
    \vspace{0.1cm}
    \resizebox{1.0\linewidth}{!}{
    \begin{tabular}{lcc|cc}
        \toprule
        \multirow{2}{*}{Model}   & \multirow{2}{*}{\#Head} & \multirow{2}{*}{\#Layer}  & \textit{test-} & \textit{test-} \\
                              &   &     & \textit{clean} & \textit{other} \\
        \midrule
        Trans($d=328$)        & 2  & 2   & 85.0    & 76.9 \\  % 5,093,022
        Trans($d=328$)        & 4  & 2   & 85.9    & 77.8 \\  % 5,093,022
        Trans($d=328$)        & 8  & 2   & 86.9    & 79.1 \\  % 5,093,022
        Trans($d=320$)        & 16 & 2   & 87.2    & 79.4 \\  % 4,916,710
        Trans($d=248$)        & 4  & 4   & 88.4    & 81.4 \\  % 5,086,206
        \midrule
        Conf($d=256, k=17$)        & 2  & 2   & 88.8    & 82.1 \\  % 5.084,198
        Conf($d=256, k=17$)        & 4  & 2   & 88.9    & 82.2 \\  % 5.084,198
        Conf($d=256, k=17$)        & 8  & 2   & 89.2    & 82.6 \\  % 5.084,198
        Conf($d=256, k=17$)        & 16 & 2   & 89.3    & 82.7 \\  % 5.084,198
        Conf($d=192, k=9$)         & 4  & 4   & 90.3    & 84.4 \\  % 5,153,382
        \bottomrule
    \end{tabular}}
    \label{tab:sa_ablation}
    \vspace{0.2cm}
\end{table}

Table~\ref{tab:sa_ablation} presents the effect of the number of attention heads in phoneme recognition.
We observe that accuracy consistently increases as the number of heads increases.
We assume that the improvement is because more diverse phonetic relationships can be captured from one layer~\cite{understanding}, especially for the challenging \textit{test-other} dataset.
The results also show that increasing the number of layers (4 heads, 4 layers) achieves higher accuracy than increasing the number of heads (8/16 heads, 2 layers).
% While previous automatic speech recognition models~\cite{conformer,trans_transducer} utilize the same number of heads (e.g., 4 or 8) for every self-attention layer, the results imply that using the different number of heads for different layers may further improve the performance, as suggested in previous work~\cite{similarity_content}.

% ---------------------------------------------------------------- %
\subsection{Transferability of Restricted Attention Range}
% ---------------------------------------------------------------- %

For Transformer and Conformer models, we can think of using different self-attention ranges for training and inference.
In other words, is the attention range restriction generally transferable?
For example, the model may access only 10 seconds of receptive field length during training but observe an unlimited range during inference.

% \begin{table}[t]
%     \centering
%     \caption{Transferability of attention range restriction through different training and inference setting.}
%     \vspace{-0.3cm}
%     \begin{subtable}[t]{1.0\linewidth}
%         \centering
%         \vspace{0.1cm}
%         \caption{\textit{test-clean} accuracy (\%)}
%         \begin{tabular}{l|cccc}
%             \toprule
%             Train\textbackslash Inference & $r=8$ & $r=24$ & $r=56$ & $r=\infty$ \\ 
%             \midrule
%             Conf($r=8$)       & 88.3 & 87.7 & 87.1 & 86.4  \\
%             Conf($r=24$)      & 88.0 & 88.4 & 88.4 & 88.1  \\
%             Conf($r=56$)      & 87.4 & 88.3 & 88.6 & 88.5  \\
%             Conf($r=\infty$)  & 79.2 & 83.2 & 85.3 & 88.9  \\
%             \bottomrule
%         \end{tabular}
%     \end{subtable}
%     \begin{subtable}[t]{1.0\linewidth}
%         \centering
%         \vspace{0.1cm}
%         \caption{\textit{test-other} accuracy (\%)}
%         \begin{tabular}{l|cccc}
%             \toprule
%             Train\textbackslash Inference & $r=8$ & $r=24$ & $r=56$ & $r=\infty$ \\ 
%             \midrule
%             Conf($r=8$)       & 81.1 & 80.2 & 79.4 & 78.5  \\
%             Conf($r=24$)      & 80.7 & 81.4 & 81.4 & 81.0  \\
%             Conf($r=56$)      & 80.0 & 81.3 & 81.7 & 81.7  \\
%             Conf($r=\infty$)  & 71.1 & 76.4 & 78.9 & 82.2  \\
%             \bottomrule
%         \end{tabular}
%     \end{subtable}
%     \vspace{0.1cm}
%     \label{tab:transfer}
% \end{table}

\begin{table}[t]
    \centering
    \caption{Transferability of attention range restriction through different training and inference setting.}
    \vspace{-0.3cm}
    \begin{subtable}[t]{1.0\linewidth}
        \centering
        \vspace{0.1cm}
        \caption{\textit{test-clean} accuracy (\%)}
        \begin{tabular}{l|cccc}
            \toprule
            Train\textbackslash Inference & $r=12$ & $r=28$ & $r=60$ & unlimited \\ 
            \midrule
            Conf($r=12$)      & 89.5 & 89.0 & 88.3 & 87.7  \\
            Conf($r=28$)      & 89.4 & 89.7 & 89.7 & 89.4  \\
            Conf($r=60$)      & 89.0 & 89.6 & 89.8 & 89.7  \\
            Conf(unlimited)   & 87.4 & 89.0 & 89.6 & \textbf{90.3}  \\  % CAN WE DYNAMICALLY CHANGE THE RECEPTIVE FIELD LENGTH TO ADAPT?
            \bottomrule
        \end{tabular}
    \end{subtable}
    \begin{subtable}[t]{1.0\linewidth}
        \centering
        \vspace{0.2cm}
        \caption{\textit{test-other} accuracy (\%)}
        \begin{tabular}{l|cccc}
            \toprule
            Train\textbackslash Inference & $r=12$ & $r=28$ & $r=60$ & unlimited \\ 
            \midrule
            Conf($r=12$)      & 83.0 & 82.2 & 81.4 & 80.7  \\
            Conf($r=28$)      & 83.1 & 83.5 & 83.5 & 83.1  \\
            Conf($r=60$)      & 82.5 & 83.4 & 83.8 & 83.7  \\
            Conf(unlimited)   & 79.8 & 82.6 & 83.6 & \textbf{84.4}  \\  % DRAMATIC DECREASE
            \bottomrule
        \end{tabular}
    \end{subtable}
    % \vspace{0.1cm}
    \label{tab:transfer}
\end{table}

Table~\ref{tab:transfer} shows the accuracy of Conformer models that exploit different attention range restrictions for training and inference.
The model configuration follows Table~\ref{tab:receptive_field}.
In short, the model best performs with the same range restriction used during training.
The performance degradation grows as the difference between $r$ used in training and inference increases.
Furthermore, the model trained with an unlimited attention range suffers the most from restricting the attention range.
We assume this is because such model is trained to exploit global information more than the local characteristics.

% ---------------------------------------------------------------- %
\subsection{Inference Speed}\label{ssec:speed}
% ---------------------------------------------------------------- %

The number of parameters is an important factor for practical usage, but the inference speed does not directly correspond to the parameter size.
We measure the GPU inference speed of different architectures and find out in which situation each architecture is advantageous.

% \begin{table}[t]
%     \centering
%     \caption{Inference speed comparison on GPU, reported in a milliseconds-per-sequence.}
%     \vspace{0.2cm}
%     \begin{tabular}{c|ccccccc}
%         \toprule
%         \multirow{2}{*}{Model} & \multicolumn{7}{c}{Sequence length(sec)} \\ 
%                         & 5 & 10 & 15 & 20 & 25 & 30 & 35 \\  % 128, 248, 376, 496, 624, 736, 872
%         \midrule
%         CNN             & 0.29 & 0.32 & 0.35 & 0.38 & 0.42 & 0.47 & 0.52 \\ 
%         RNN             & 0.22 & 0.43 & 0.61 & 0.84 & 0.99 & 1.19 & 1.45 \\ 
%         Trans           & 0.25 & 0.36 & 0.50 & 0.72 & 1.00 & 1.33 & 1.75 \\ 
%         Conf            & 0.30 & 0.42 & 0.58 & 0.80 & 1.07 & 1.35 & 1.77 \\ 
%         \bottomrule
%     \end{tabular}
%     \label{tab:speed}
%     % \vspace{0.1cm}
% \end{table}

\begin{figure}[t]
    \centering
    \includegraphics[width=1.0\linewidth]{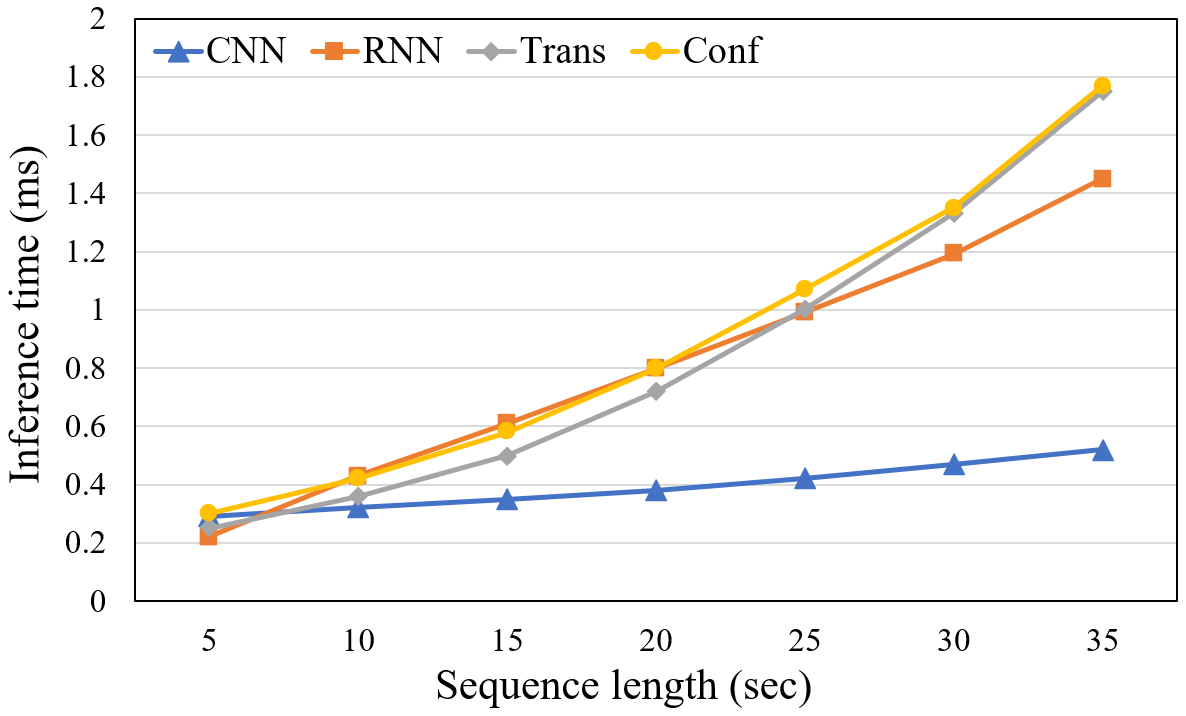}
    \caption{Inference speed comparison on GPU. The results are reported in a milliseconds-per-sequence.}
    \label{fig:speed}
    % \vspace{-0.1cm}
\end{figure}

Figure~\ref{fig:speed} shows the inference speed of the encoder part, the middle block in Figure~\ref{fig:architecture}, estimated on a single A100 GPU.
The models in Figure~\ref{fig:speed} are the best models in Table~\ref{tab:receptive_field}.
Note that the parameter size of models is almost the same (5M).
During inference, we fix the batch size to 64 and only change the input sequence length.
We observe that ContextNet and LSTM models show linearly increasing inference time of $O(T)$, while Transformer and Conformer show quadratically increasing time of $O(T^2)$.
Especially, ContextNet is significantly faster when the input sequence length is longer than 10 seconds.
We assume this is mainly due to the efficient DS convolution.
Conformer is slightly slower than Transformer because of an additional convolution module; however, as the sequence length increases, the gap decreases as the proportion of self-attention computation increases.

%% file: sections/conclusion.tex
% ---------------------------------------------------------------- %
\section{Conclusion}\label{sec:conclusion}
% ---------------------------------------------------------------- %

We presented a comparative study of various DNN architectures through the lens of phoneme classification. 
We compared CNN, RNN, Transformer, and Conformer models from three different perspectives.
The performance of CNN saturates when the receptive field length exceeds about 5 seconds, but Transformer and Conformer continue to increase in performance.
This is because they employ the self-attention mechanism to cluster and utilize similar phonemes at a distance. 
When the parameter size allowed is small, such as 1M, the ContextNet performs best.
In addition, the ContextNet is very advantageous in inference speed.

% We presented a comparative study of various DNN architectures through the lens of phoneme classification.
% We compared CNN, RNN, Transformer, and Conformer models from three different perspectives.
% In particular, we empirically demonstrate the advantage of self-attention in extracting long-range phonetic relationships.